\documentclass[letterpaper,10pt]{article}
\usepackage[top=30mm, bottom=30mm, left=15mm, right=15mm]{geometry}
\usepackage{amsmath}
\usepackage{amssymb,latexsym}
\usepackage[font={small,it}]{caption}
\usepackage[T1]{fontenc}
\usepackage[utf8]{inputenc}
\usepackage{graphicx}
\usepackage{bm}
\usepackage{array}
\usepackage{dcolumn}
\usepackage{multicol}
\usepackage{hhline}
\usepackage{epsfig}
\usepackage{amsmath}
\usepackage{amssymb}
\usepackage{lmodern}
\usepackage{xcolor}
\usepackage{tablefootnote}
\usepackage{makecell}

\newcolumntype{C}[1]{>{\centering\arraybackslash}m{\dimexpr#1-2\tabcolsep\relax}}
\colorlet{linkequation}{blue}
\usepackage[colorlinks]{hyperref}

\newcommand{\be}{\begin{equation}}
	\newcommand{\ee}{\end{equation}}
\newcommand{\bea}{\begin{eqnarray}}
	\newcommand{\eea}{\end{eqnarray}}
\makeatletter
\newcommand*{\rom}[1]{\expandafter\@slowromancap\romannumeral #1@}
\makeatother
\begin{document}
	\title{Weak Energy Condition, Trapped Surfaces and Black hole Third Law}
	\author{F. Bahmani$^{1}$, F. Shojai$^{1}\footnote{Corresponding author: fshojai@ut.ac.ir}$\ and Sh. Anjomshoaa$^{1}$\\$^1$Department of Physics, University of Tehran, P.O. Box 14395-547\\Tehran, Iran.}
	\maketitle
		\begin{abstract}
	We consider the third law of thermodynamics for families of 4- and n-dimensional Vaidya black holes, including many of interest. 
Since there are several versions of the definition of surface gravity as well as the extremality condition for dynamical black holes, we first show that for the considered 4- and n-dimensional Vaidya families, these definitions are consistent with each other. 
We assume, first, that a non-extremal black hole evolves to an extremality state after a finite time and second,  that the weak energy condition for the source holds at all times. 
We then compare the results of these assumptions and investigate whether there are ranges of black hole parameters where these two assumptions are in conflict.
				\end{abstract}
	\section{Introduction}
  The laws of ‌black hole (BH) thermodynamics were proposed by Bardeen, Carter and Hawking based on the remarkable similarities  between the laws of thermodynamics and certain properties of BHs \cite{Bardeen:1973gs}.  This was further strengthened by the discovery of Hawking radiation \cite{Hawking:1974rv}, which introduced the concept of BH temperature.
	
The third law of thermodynamics which is of interest here, accepts two different formulations, both are due to Nernst \cite{ner}. The first one is called the "entropic formula". It states that the entropy of a system approaches a constant as the temperature approaches absolute zero .  This constant can be set equal to zero according to Planck's hypothesis \cite{plank}. The entropic formulation of the third law is clearly violated by Kerr-Newman BHs \cite{Does the third}.

The second formula is the weak form of the third law introduced by Israel. It is based on Nernst's theorem. It states that the surface gravity (SG) of a BH cannot be reduced to zero by any continuous and finite-time process if the stress-energy tensor is finite and satisfies the weak energy condition (WEC) at least in the vicinity of the apparent horizon (AH) \cite{Israel:1986gqz}.
 The extremality condition for stationary BHs is expressed as the Killing SG is zero and there is no trapped
surface. These properties do not necessarily coincide for the dynamical BHs, and even other definitions of extremality can be considered, such as the coordinate-invariant definition proposed in \cite{Pielahn:2011ra}.
 
 For dynamical BHs, there are several quasi-local horizons \cite{Booth}. The most important one is the AH which is not necessarily a null surface. It is not generated by some geodesics and its existence depends on the space-time foliation \cite{Wald:1991zz}. Consequently, it should be clear which evolving horizon should appear in the thermodynamic laws of the BH. For example, in the second law, for which horizon the area is a non-decreasing function with time? We know that even for spherically symmetric  BHs, the area of a evolving horizon can increase or decrease depending on its nature \cite{Nielsen:2008cr}. A similar question arises for  the entropy of the dynamical BH. In situations where the BH is evolving by processes such as accretion, radiation, or embedding in a time-dependent background, the area of which horizon gives the entropy of the BH?  Is it  related to the event horizon or to the marginally trapped surface?
Recently, it has been shown that the radiation temperature of the multi-horizon BH is not the same as the conventional Hawking temperature defined for the outer event horizon, even for the stationary case without considering back-reaction effects \cite{Singha:2022uny}. In fact, the contribution between the horizons of a multi-horizon spacetime determines the Hawking temperature. The universal thermodynamics for dynamical BHs and their radiation as a tunneling effect have been analyzed in \cite{DiCriscienzo:2007pcr,Bhattacharjee:2018kro,Firouzjaee:2011hi}. Correspondingly, which prescription of the SG should appear in the first law, since there are different prescriptions of the SG for a non-Killing horizon \cite{faraoni}. It is known that if we define the SG as the degree of non-affinity of the null vector field perpendicular to the horizon, the SG value depends on the choice of normalization of the null vector \cite{Pielahn:2011ra}. Therefore, no unique SG is defined.

Another natural question is whether the extremal BHs can be produced by gravitational collapse? If so, is the required time scale proportional to the inverse of the SG? It is known that as a stationary BH approaches the extremality, some of its physical properties can change discontinuously \cite{Does the third}. Thus, the extremal BH is not necessarily the limit of a near extremal BH. This limit is also questionable from a thermodynamic point of view. According to the quantum field theory in curved spacetime, the extremal BHs cannot be thermal objects. Therefore, the concept of zero temperature is unclear for them \cite{QF T}. 
As a result, a semi-classical theory of gravity, including back-reaction effects, may be required to understand the third law of BH thermodynamics. 
In \cite{Anderson:2000pg} it is shown that macroscopic zero temperature BHs do not exist. Any static zero temperature semiclassical BH must then be microscopic and not smoothly matched to the classical extremal BH.
 The test body approximation, in which a test body is absorbed by the BH, clearly does not take into account the concept of finite advanced time used in Israel's theorem \cite{test}.  
In fact, only the arrival time of the particle at the horizon surface is considered in this approximation. 
The time taken by the BH to absorb the energy and angular momentum of the particle is the time scale considered in the formulation of Israel's third law.   
 In the test body approximation, the absorption is ideally instantaneous. However,  the test body has a finite size. Therefore,  it actually takes a finite time to cross the horizon.

Furthermore, under the assumption of local equilibrium conditions and using the Bekenstein bound \cite{Bekenstein:1974jk}, it is shown that the lower bound for the relaxation time of the thermal system is proportional to the inverse of the temperature or SG \cite{test}. Therefore a very long time is required for the absorption of the test object in the final stage of collapse of a near-extremal BH. In addition, it can be shown that the absorption of a test body by a near-extremal BH cannot be described as a sequence of equilibrium states. In fact, this process can only be adiabatic if it is infinitely slow. Thus, absorption in finite time is not equivalent to an equilibrium process. In this sense, even if the test object can reach the horizon in finite time, an extremal BH cannot be generated in finite time in a continuous process \cite{test}. Therefore, the third law of BH thermodynamics is not really violated. 

A counterexample of the third law in n-dimensional (n-D) Einstein-Gauss-Bonnet gravity \cite{EG} with a negative cosmological constant can be found in \cite{Torii}. In the "non-general relativity" branch of this theory, the motion of a thin shell with a mass equal to the mass of the extremal BH in 6D space-time is studied \cite{Torii}. Using the equation of motion for a thin dust shell and generalized Israel's junction conditions, the shell collapses from infinity toward the center. This is because the effective potential is negative everywhere and thus, the extremal horizon is formed.

There are other issues raised by Israel who associated a disconnected outermost AH with a singularity of the space-time metric or matter \cite{Kehle:2022uvc}. Farrugia et al. \cite{yy} showed that if a charged singular massive shell is thrown into a non extremal Reissner-Nordstr\" om BH, under certain conditions on the injection energy, the ratio of the initial velocity and charge of the thin shell to the mass and charge of the BH, the BH is transformed into an extremal one, while the outermost AH is disconnected. 
Thus, some process leading to the discontinuity of the AH, may be related to the violation of the third law, according to \cite{yy}. By explicitly requiring that the stress-energy tensor of the accreted matter remains bounded in a neighborhood of the AH, Israel dismissed this counterexample in his formulation of the third law.
Using a novel gluing procedure, it has been shown \cite{Kehle:2022uvc} that a BH can be formed from regular, one-ended, asymptotically flat Cauchy data in the context of the spherically symmetric Einstein-Maxwell charged scalar field gravity. The mentioned  BH collapses gravitationally and forms an exactly Schwarzschild AH, while at a later time there is an exactly extremal Reissner-Nordstr\"om event horizon. Thus, subextremal BHs can become extremal in finite time. 
Despite the discontinuity of the AH, this example is regular and the scalar field matter manifestly obeys the dominant energy condition.
The authors of  \cite{our1}, \cite{our2} have analytically studied the Oppenheimer–Snyder collapse  of a star into a special class of regular and 4D-EGB \cite{Glav} BHs.  According to their model the stellar matter violates only the strong energy condition  and the radius of the star reaches the extremal horizon in a finite time, and the outer SG becomes zero.

In this paper, we will investigate whether the third law is essentially a result of the WEC. To do this, we use some of the 4-D  and n-D Vaidya families of BH solutions. These contain most of the known dynamical BH solutions. 
We assume that the BH evolves to its extremal state, i.e. zero SG, in finite time and that the WEC is satisfied for the matter source.
We then compare the result with the WEC condition. 
In this way we check whether we can obtain an extremal BH by adjusting the free parameters of the metric without violating the WEC or not.
 
This paper is organized as follows. 
Two prescriptions of dynamical SG, Fodor's and Hayward's SG, and their properties are presented in section \ref{sec2}. 
Following \cite{Poisson}, section \ref{sec3} is devoted to showing that the WEC is the sufficient condition for the third law for the Vaidya Reissner-Nordstr\"om BH.
After introducing some families of Vaidya BHs in section \ref{sec4},  we analytically investigate whether the WEC assumption can guarantee that the third law is satisfied or not.
This is done in section \ref{sec5}. In Section \ref{secpre6}, we extend our approach to the study of the third law of thermodynamics for multi-horizon BHs. For some specific BHs, the results are summarized  and the main points of our approach form the content of section \ref{sec6}.

Throughout this paper, the signature of the metric tensor is assumed to be (-, +, +, +). We use geometric units, i.e., $G = c = 1$ and a dot (a prime) over a variable denotes its derivative with respect to the advanced time (radial coordinate).
	\section{Surface gravity}\label{sec2}
	In stationary space-time, the SG is a measure of the Killing vector inaffinity  at the Killing horizon, where the Killing vector becomes null. 
The SG of a non-Killing horizon in a spherically symmetric BH is usually defined as the inaffinity of the outgoing null vector $\ell^{\alpha}$ at the horizon, where the integral curve of $\ell^{\alpha}$ is
geodesic. Several  proposals have been made to fix the SG. To study the third law of thermodynamics, we will focus on two approaches to finding the SG. 
	 There are other definitions that we will not discuss here. This is because they do not give the correct value of the SG in stationary space-time or they use a special normalization of the outgoing null vector after choosing the coordinate system \cite{vissern}. A general overview of different definitions of SG can be found in \cite{Hayward1994} and
\cite{Abreu:2010ru}.

	The first one is the proposal of Fodor \cite{Fodor:1996rf}. In Fodor’s method, space-time is a foliation of null surfaces on which the  ingoing null Eddington–Finkelstein coordinate $v$ is constant and thus ingoing null rays are affinely parametrized geodesics.
	Our requirement is that this coordinate is fixed by the proper time of some stationary observer at infinity.  
 This requires  that the space-time admits an asymptotically flat spatial infinity. It should be noted that the outgoing radial null curves are always geodesic due to the spherical symmetry and they have zero expansion at the AH. In advanced Eddington- Finkelstein coordinates, the outgoing null geodesic is uniquely determined by the condition
 $\ell^{\alpha}v_{;\alpha}=1$,  but  this is not the case in the Painlev\'e-Gullstrand coordinates\footnote{In general, we can use null vectors with arbitrary normalization coefficients. In Fodor's approach, the condition that the ingoing null rays are affinely parameterized geodesics yields a first-order partial differential equation in Painlev\'e-Gullstrand coordinates. By imposing the condition $\xi^{\alpha}v_{;\alpha}=1$  that the asymptotic time-translational Killing vector, $\xi^{\alpha}$, should be satisfied, one can fix one of the integration constants arising from the above mentioned equation, but another constant remains.}
 According to Fodor's definition  
\begin{align}\label{KF1}
\ell^{\alpha}\ell_{;\alpha}^{\beta}=\kappa_{\mathrm{F}}\ell^{\beta}
\end{align}
The 4-D general metric of spherically symmetric space-time in Eddington coordinates is 
\begin{equation}\label{el1}
	ds^2=-A(v , r)^2 f(v,r) dv^2 + 2\epsilon A(v, r) dv dr + r^2 d\Omega^2
	\end{equation}
where $A(v, r)$ and $f (v, r)$ are arbitrary functions. For an ingoing (outgoing) radial flow, $\epsilon=1$ ($\epsilon=-1$) and $v$ represents the advanced (retarded) time. From \eqref{KF1}, using the Christoffel symbols of metric \eqref{el1},   we have
\begin{align}\label{KF2}
\kappa_{\mathrm{F}}=\Gamma^{0}_{00}=\frac{\dot A}{A}+\frac{(A^2f)'}{2A}
\end{align}
with the AH  at $f(v,r_{ap})=0$\footnote{A simple calculation shows that the expansion parameter of the outgoing null geodesics is $Af /r$. Thus, at $f = 0$, the expansion parameter is zero for $A = 1$.}.

The second one is proposed by Hayward \cite{Hayward:1997jp}. It is based on defining the SG of spherically symmetric space-time in terms of the Kodama vector $K^{\alpha}\equiv \epsilon^{\alpha \beta } \nabla_\beta {\cal R}$ where $\cal R$ is the areal radius of the space-time metric. The Kodama vector satisfies\footnote{ The 2-volume form $\epsilon ^{\beta \mu}$  appears on the right hand side of the Kodama equation. It is omitted in \cite{nielsend}, \cite{faraoni} and other literature. 
To see this point, it is necessary to compute 
\begin{align*}
\frac{1}{2}g^{\alpha \beta}&K^{\mu}(\nabla_{\mu}K_{\alpha}- \nabla_{\alpha}K_{\mu})\nonumber\\
&=\frac{1}{2}g^{\alpha \beta}\nabla_{\nu}{\cal R}(\epsilon ^{\mu \nu}\epsilon _{\alpha \gamma}\nabla_{\mu}\nabla^{\gamma}{\cal R}-\epsilon ^{\mu \nu}\epsilon _{\mu \gamma}\nabla_{\alpha }\nabla^{\gamma}{\cal R})\nonumber \\&=
 \frac{1}{2}(\nabla^{\beta}{\cal R})\square {\cal R}\nonumber\\
&\equiv \kappa_{_\mathrm{H}}\epsilon ^{\beta \mu}\epsilon _{\mu \nu}\nabla^{\nu}{\cal R}
= \kappa_{_\mathrm{H}}\epsilon ^{\beta \mu}K_{\mu}
\end{align*}
Note that the Kodama vector and the corresponding SG are not affected by this correction.}
\begin{align}\label{HoH}
\frac{1}{2}g^{\alpha \beta}& K^{\mu}(\nabla_{\mu}K_{\alpha}- \nabla_{\alpha}K_{\mu})= \kappa_{_\mathrm{H}}\epsilon ^{\beta \mu}K_{\mu}
\end{align}
and gives the Hayward SG as 
\begin{align}\label{KH}
 \kappa_{\mathrm{H}}=\frac{1}{2} \Box_{(h)}\cal R
 \end{align}
where $h_{\mu \nu}$ is the two-metric of $(v, r)$ space. Thus, the Hayward SG of the metric \eqref{el1} is 
\begin{equation}
\label{KH2}
\kappa_{\mathrm{H}} = \frac{(Af)'}{2A}
 \end{equation}
which  obviously corresponds to Fodor's definition of SG \eqref{KF2} for the case $A = 1$. The extremality condition for stationary BHs is that the Killing SG must be zero. This means  that $f'(r)=f(r)=0$ has a double root and for a quadratic function $f(r)$ in the variable 
$1/r$, there is no trapped surface.
However, for a function $f(r)$ of higher order, besides an extremal horizon,  BH can have a trapped region between other horizons according to the behavior of the function $f(r)$.
 \\
For dynamical BHs, according to \eqref{KF2} and \eqref{KH2}, the condition of zero SG does not necessarily lead to satisfying the conditions $f'(v,r)=f(v,r)=0$. These conditions mean that the two roots of the function $f(v,r)$ coincide with each other and the BH has an extermal AH.
In other words, the condition of zero SG in general dynamical situations does not imply the elimination of the trapped region between two horizons, even for metrics with a quadratic function $f(v,r)$  in the variable 
$1/r$.
A coordinate invariant definition of extremality for the evolving horizon, is given in \cite{Pielahn:2011ra} as 
\begin{align}\label{ex}
n^{\alpha}\nabla_{\alpha}\theta_{\ell}=0
\end{align}
where $\theta_{\ell}$ is the expansion parameter of the outgoing null geodesics and $n^{\alpha}$ is the ingoing radial null ray of metric \eqref{el1}. A simple calculation leads to  
\begin{align}\label{ex1}
 n^{\alpha}\nabla_{\alpha}\theta_{\ell} = \frac{Af-rfA'-rAf'}{Ar^2}
\end{align}
Thus, the extremality condition \eqref{ex} corresponds to the zero value of Fodor's SG \eqref{KF2} and Hayward's SG \eqref{KH2} at the horizon, only for the case $A = 1$\footnote{If $A=1$, the zero value of Fodor's SG \eqref{KF2} and Hayward's SG \eqref{KH2} at the AH, $(f(v,r_{ap})=0)$, leads to the condition $f'(v,r_{ap})=0$. Also the condition \eqref{ex} according to \eqref{ex1}, by assuming $A=1$ at AH, leads to  $f'(v,r_{ap})=0$ which is compatible with the zero value of Fodor's SG \eqref{KF2} and Hayward's SG \eqref{KH2}.}.
\section{Third law of BH thermodynamics}\label{sec3}
The third law of BH mechanics states that if the stress-energy tensor is bounded and satisfies the WEC in a neighborhood of the AH, then the SG of a BH cannot be reduced to zero within a finite advanced time. Werner Israel in \cite{Israel:1986gqz} gave a precise formulation of this law.  
In the following sections, we will illustrate this law with some examples. We want to know whether the imposition of the WEC in the AH prevents the SG of the horizon from being zero.

For the simplest example \cite{Poisson},  consider a charged BH given in the ingoing Vaidya coordinate \eqref{el1} with
$f=1-2m(v)/r+q^2(v)/r^2$
where $m(v)$ and $q(v)$ are the time-dependent mass and charge of the BH. This is a solution of Einstein’s equations given by the stress-energy tensor 
$T^{\alpha \beta}=\rho \ell^{\alpha}\ell^{\beta}+T_{EM}^{\alpha \beta}$
where the first term comes from the null dust with the density 
$\rho=(\dot m-q \dot q/r)/(4\pi r^2)$
and the second is the contribution of the electromagnetic field 
$T^{\alpha}_{ \beta}=P diag(-1,-1,1,1)$ where $P=q^2/(8\pi r^4)$.
This means that the energy density measured by an observer with four velocities $u^{\alpha}$ will always be positive. If we restrict  ourselves to the radial  observer, $T^{\alpha \beta}u^{\alpha }u^{\beta}=\rho(dv/d\tau)^{2}+P$. 
In the limit of small values of $dv/d\tau$, the WEC is obviously satisfied. However, for sufficiently large values of $dv/d\tau$, the WEC gives $\rho>0$. Evaluating this on the event horizon $r_+=m+\sqrt{m^2-q^2}$ gives $m \dot m-q \dot q+\sqrt{m^2-q^2}\dot m>0$.
Now, if the BH becomes extremal at advanced time $v_0$,  this means that $\Delta(v_0)=m(v_0)-q(v_0)=0$. Before $v_0$ the BH is not extremal so $\Delta(v)>0$ for $v<v_0$ and so the  function $\Delta(v)$ decreases as  $v$ approaches $v_0$. This contradicts with the result of WEC which gives $m(v_0)\dot{\Delta}(v_0)>0$ and so  $\Delta(v)$ is an increasing function. We see that WEC prevents the BH from becoming extremal at a finite advanced time. 

In the following, this method is used for some general Vaidya BHs that have an extremal horizon.
By adjusting the free parameters of the metric, we want to see if we can obtain an extremal BH in finite time without violating the WEC.
	\section{General Vaidya BHs}\label{sec4}
		In this section, we have a review on the spherically symmetric solutions of Einstein’s equations describing general classes  of Vaidya BHs.
			\subsection{A family of 4-D Vaydia BHs}\label{sec4.1}
 	Kothawala et al. \cite{Kothawala:2004fy} have proved a theorem that can be easily used to generate dynamical solutions of BHs in general relativity, by imposing some conditions on the stress-energy tensor. This theorem is a generalization of the earlier theorem of Salgado in \cite{Salgado}
 	who proved a theorem characterizing
a three parameter family of static and spherically
symmetric solutions to Einstein's equations
by imposing certain conditions on the stress-energy tensor. This theorem was generalized for n-D by Gallo \cite{gallo}. 
	 	  The authors of \cite{Kothawala:2004fy} considered the line element \eqref{el1} as a solution of 
	 	  Einstein's equations with a null fluid source (radiation) whose stress-energy tensor satisfies $T_r^v =0$ and	$T_{\theta}^{\theta}= k T_r ^r$ ($k$ is an arbitrary real number). Consider the special case $T_r^v =0$ which implies that  $A(v, r) = g(v)$ due to Einstein's field equations. By introducing the new null coordinate $\bar v = \int g(v) dv$, we can always set $A(v, r) = 1$ without loss of generality. This  hypothesis also implies that $G_r^r =G_v^v$ i.e., $T_r^r =T_v^v$ . Here, we set $\epsilon = 1$. Thus, the  coordinate $v$ represents the advanced Eddington time. It is useful to introduce a local mass function
$m(v, r)$ defined by $f(v, r)=1-2m(v, r)/r$. By enforcing the conservation law of the stress-energy tensor and the above assumptions on its components, the diagonal elements of the stress-energy tensor are obtained as
	\begin{equation}\label{hus2}
	T_{\beta}^{\alpha}=\frac{C(v)}{r^{2(1-k)}} diag[1, 1, k, k]
		\end{equation}
where $C(v)$ is an arbitrary function. Now, the diagonal components of Einstein's equations can be easily integrated to give 
\begin{equation}\label{eq:L}
		m(v, r)=
		\begin{cases}
						M(v)-\frac{4\pi C(v)}{2k+1} r^{2k+1} & k\ne-\frac{1}{2} \\
			M(v)-4\pi C(v)\ln {r} & k=-\frac{1}{2} 
		\end{cases}
	\end{equation} 	
	Here the arbitrary function $M(v)$ is obtained by integration. The required non-zero off-diagonal component $T_v^r$ is then given by Einstein's equations as
		\begin{equation}\label{T12}
		T^r_v=
		\begin{cases} 
			\frac{1}{4\pi r^2} \dot{M}(v)-\frac{1}{2k+1} \dot{C}(v) r^{2k-1} & k\ne -\frac{1}{2} \\
			\frac{1}{4\pi r^2} \dot{M}(v)-\frac{1}{r^2} \dot{C}(v) \ln r  & k=-\frac{1}{2}
		\end{cases}
	\end{equation}
	Including the cosmological constant $\Lambda$, a generalization of the above theorem gives \cite{Kothawala:2004fy}
	\begin{equation}
	m(v, r)=
		\begin{cases}
						M(v)-\frac{4\pi C(v)}{2k+1} r^{2k+1}+\frac{\Lambda r^3}{6} & k\ne-\frac{1}{2} \\
			M(v)-4\pi C(v)\ln {r}+\frac{\Lambda r^3}{6} & k=-\frac{1}{2} 
		\end{cases}
	\end{equation}
 In this case, the diagonal elements of the stress energy tensor are given by
	\begin{equation}\label{hus22}
	T_{\beta}^{\alpha}=\frac{C(v)}{r^{2(1-k)}} diag[1, 1, k, k]-\frac{\Lambda}{8\pi} diag[1, 1, 1, 1]
		\end{equation}
		and $T^r_v$ has the same form as \eqref{T12}.
		What was mentioned above is the content of the theorem mentioned in \cite{Kothawala:2004fy}.
	By choosing arbitrary functions $M(v)$ and $C(v)$, many BH solutions can be generated such as the Vaidya  \cite{vady,vady2} , Bonnor- Vaidya \cite{bonor6}, dS/AdS \cite{w10}, global monopole \cite{M19}, Husain \cite{Hos7} and Kiselev \cite{kiselv} BHs. We refer the interested reader to the Table 1 in \cite{Kothawala:2004fy}.
 For Bonnor- Vaidya BH, 
	\begin{equation}\label{bon1}
 C(v)=-\frac{q(v)^2}{8\pi},\qquad k=-1
	\end{equation}
		\begin{equation}\label{bon2}
		T_{\beta}^{\alpha}=-\frac{q(v)^2}{8\pi r^4} diag[1, 1, -1, -1]\qquad
		T_v^r=\frac{1}{4\pi r^3}[r\dot{M}(v)-q(v)\dot{q}(v)]
	\end{equation}
	for Husain BH,
		\begin{equation}\label{hus1}
	C(v)=-\frac{g(v)}{4\pi},\qquad k=-m
	\end{equation}
	\begin{equation}\label{hus2}
	T_{\beta}^{\alpha}=-\frac{g(v)}{4\pi r^{2(m+1)}} diag[1, 1, -m, -m]\qquad
		T_v^r=\frac{1}{4\pi r^2}[\dot{M}(v)-\frac{1}{(2m-1) r^{(2m-1)}}\dot{g}(v)]
	\end{equation}
 and for Kiselev BH\footnote{It is a static spherically symmetric solution of Einstein equations with quintessential matter. The Kiselev BH is well motivated by the fact that BHs in the real world are not isolated  and are not part of an empty background.}
 	\begin{equation}\label{hus1}
 C(v)=\frac{3wN(v)}{8\pi},\qquad k=-\frac{(3w+1)}{2}
	\end{equation}
	\begin{equation}
	T_{\beta}^{\alpha}=\frac{3w N(v) r^{-3(1+w)}}{8\pi} diag[1, 1, -\frac{1+3w}{2},-\frac{1+3w}{2}]\label{tk}\qquad
		T_v^r=\frac{2\dot{M}(v) +r^{-3w} \dot{N}(v)}{8\pi r^2}
	\end{equation}
		 	where $m$ is a constant and $w$ is the parameter of the equation of state of quintessential matter.
		\subsection{A family of 4-D regular Vaidya BHs}
	Here we consider a family of  4-D regular BHs whose the local mass function is not in the form of \eqref{eq:L}. 
		It is well known that in order 
	to get rid of singularities inside the BHs, one can modify the Einstein–Hilbert action by additional terms arising from vacuum polarization and particle creation \cite{Mukhanov}, or assume that the curvature is bounded by some fundamental value \cite{Markov,Markov2,Markov3,Frolov2}. One of the most important steps to remove the singularity was taken by Bardeen. He coupled the action of general relativity to nonlinear electrodynamics \cite{bar}.  Then, Hayward introduced a regular BH solution
with a fundamental length scale that prevents the formation of a singularity \cite{hay}. Both the Bardeen and Hayward metrics  reduce to the de Sitter metric at small distances while asymptotically they reduce to the Schwarzschild geometry. Recently, the authors of \cite{our1} consider a special class of  non-stationary spherically symmetric and asymptotically flat regular BH solutions as follows
\begin{align}\label{REG1}
ds^2&=-\left(1-\frac{2m(v)}{r\left(1+\beta(v)r^{\frac{-3}{n}}\right)^n}\right)dv^2+2dvdr+r^2d\Omega ^2
	\end{align}
This metric can be obtained from a collapsing star governed by the polytropic equation of state, $P\propto \rho^{1+1/n}$.  $m(v)$ is the geometric mass of the star with the length dimension in geometric units. $\beta(v)$ is an arbitrary dimensionless positive parameter and we assume that the polytropic index $n$ is positive. The positive function $\beta (r)$ regularizes the singularity of the Schwarzschild metric at $r = 0$  and setting $n=1,3/2$ gives the Hayward and Bardeen metrics respectively. The required non-zero  components  of the stress-energy tensor are given by
\begin{align}
T^v_v&=T^r_r=\frac{-6m(v)\beta(v)}{8\pi \left(r^{\frac{3}{n}}+\beta(v)\right)^{n+1}} \label{REG2}\\
T^r_v&=\frac{2r\left((r^{\frac{3}{n}}+\beta(v))\dot{m}(v)-nm(v)\dot{\beta}(v)\right)}{8\pi \left(r^{\frac{3}{n}}+\beta(v)\right)^{n+1}}\label{REG3}
	\end{align}
\subsection{A family of n-D Vaidya BHs}
 In \cite{Ghosh:2008zza}, Ghosh et al. extended the previous theorem that was mentioned in section \eqref{sec4.1} to the n-D  space-time 
\begin{equation}\label{sha}
		ds^2= -A(v , r)^2 f(v,r) dv^2 + 2\epsilon A(v, r) dv dr + r^2 d\Omega_{n-2} ^2
	\end{equation}
with the local mass function 
\begin{equation}\label{dn0}
		f(v, r) =
		1-\frac{ 2 m(v, r)}{(n-3)r^{(n-3)}}
	\end{equation}
	As before, without loss of generality, we can set $A(v, r) = 1$ which reduces the metric \eqref{sha} to the standard n-D Vaidya metric.  The authors of \cite{Ghosh:2008zza} use the non-vanishing components of the Einstein tensor. They assume that the stress-energy tensor satisfies in the conditions $T_r^v =0$ and $T_{\theta_{1}}^{\theta_{1}} =k T_r^r$ where $\theta_1$  is the azimuthal angle in the (n-2)D sphere. They then apply the method of the previous section  and show that the metric \eqref {sha} is a solution of Einstein's equations if
		\begin{equation}\label{dn8}
		m(v, r)=
		\begin{cases}
					M(v)-8\pi C(v) \frac{n-3}{n-2} \frac{1}{(n-2)k+1} r^{(n-2)k+1} & k\ne-\frac{1}{n-2} \\
			M(v)-8\pi C(v) \frac{n-3}{n-2} \ln(r) & k=-\frac{1}{n-2}	
		\end{cases}
	\end{equation}
		\begin{equation}\label{dn6}
			T_{\beta}^{\alpha}=\frac{C(v)}{r^{(n-2)(1-k)} } diag[1, 1, k, ..., k]
	\end{equation}
		\begin{equation}\label{dn10}
		T_v^r=
		\begin{cases}
			\frac{1}{8\pi r^{n-2}} (\frac{n-2}{n-3})  \dot{M}(v)- \frac{1}{(n-2)k+1}  \dot{C}(v) r^{(n-2)(k-1)+1} & k \ne \frac{-1}{n-2} \\
			\frac{1}{8\pi r^{n-2}} (\frac{n-2}{n-3})  \dot{M}(v)- \frac{1}{r^{n-2}}  \dot{C}(v) \ln(r) & k=-\frac{1}{n-2} 	
		\end{cases}
	\end{equation}
where $M(v)$ and $C(v)$ are arbitrary functions of $v$.
 The extensions to arbitrary dimensions of certain metrics listed at the end of the previous section also follow from \eqref {sha}-\eqref {dn8}.
These are  n-D Vaidya \cite{Dadhich28,Iyer34}, Bonnor-Vaidya \cite{Patel29,Chatterjee35}, dS/AdS \cite{Patel29}, universal unipolar \cite{Patel29,Rocha30,BarriolaM36} and Husain \cite{Patel29,Rocha30,Husain14}  metrics.
The authors of \cite{Ghosh:2008zza} then generalize the above theorem to include $\Lambda$ and obtain
\begin{equation}
m(v,r) = \left\{ \begin{array}{ll}
       M(v) - 8 \pi C(v) \left(\frac{n-3}{n-2}\right) \frac{1}{(n-2)k + 1}r^{(n-2)k + 1} + \frac{(n-3)}{(n-2)(n-1)} \Lambda r^{n-1}   &
\hspace{.1in}     \mbox{ 
 $k \neq \frac{-1}{(n-2)}$}, \\
         & \\
       M(v) - 8 \pi C(v) \left(\frac{n-3}{n-2}\right) \ln r + \frac{(n-3)}{(n-2)(n-1)} \Lambda r^{n-1}& \hspace{.1in}      \mbox{
 $k = \frac{-1}{(n-2)}$}.
                \end{array}
        \right.                         \label{eq:mvl}
\end{equation}  
\begin{equation}\label{land1}
T^{\alpha}_{\beta} =  \frac{C(v)}{r^{(n-2)(1-k)}} {\mbox{diag}}[1, 1, k,\ldots,
k] - \frac{\Lambda}{8 \pi} {\mbox{diag}}[1, 1, 1,\ldots, 1]
\end{equation}
and $ T^r_v $ has the same form as before.
Two examples of the metrics \eqref{eq:mvl} that we will consider here are the n-D Vaidya de Sitter-Reissner-Nordstr\"om  metric and the n-D Vaidya Bardeen-de Sitter metric. For the first one
	 	\begin{equation}\label{ADQ}
	f(v,r)=1-\frac{2m(v)}{r^{n-3}}+\frac{q^2(v)}{r^{2(n-3)}}-\frac{2\Lambda r^2}{(n-1)(n-2)}
	\end{equation}
 which gives
	\begin{equation}\label{k}
	M(v)=(n-3)m(v)  \hspace{1cm}  C(v)=(n-2)(3-n)q^2(v)/16\pi \hspace{1cm}k=-1
	\end{equation}
	Substitution of \eqref{k} in \eqref{dn10} and \eqref{land1} yields 
	\begin{align}\label{poi}
T^v_v=\frac{(n-2)(3-n)q^2(v)}{16 \pi r^{2(n-2)}}- \frac{\Lambda}{8\pi} \qquad	  
	T^r_v=\frac{(n-2)\dot{m}(v)}{8\pi r^{n-2}}-\frac{(n-2)(3-n)q(v)\dot{q}(v)}{8\pi \left(-(n-2)+1)\right)}r^{-2(n-2)+1 }   
	\end{align}
	For the second, the spherically symmetric Bardeen-de Sitter BH \cite{Fernando:2016ksb} and the generalized n-D Bardeen-de Sitter BH \cite{Ali:2018boy} were recently derived. The Vaidya line element of the latter is given by
\begin{align}\label{Bar}
f(v,r)=1-\frac{M(v)r^2}{(r^{n-2}+e^{n-2}(v))^{\frac{n-1}{n-2}}}-\frac{2\Lambda r^2}{(n-1)(n-2)}
\end{align}
where $e(v)$ is the nonlinear charge, $M(v)=16\pi m(v)/((n-2)\Omega_{n-2})$ and $\Omega_{n-2}=2\pi^{(n-1)/2}/\Gamma((n-1)/2))$.
The required non-zero  components  of the stress-energy tensor are given by
\begin{align}
T^r_v&=\frac{-(n-2)}{16\pi r}\left(-\frac{\dot{M}(v)r^2}{(r^{n-2}+e^{n-2}(v))^{\frac{n-1}{n-2}}}+\frac{(n-1)M(v)r^2e^{n-3}(v)\dot{e}(v)}{(r^{n-2}+e^{n-2}(v))^{\frac{2n-3}{n-2}}}\right)\\
T^v_v&=\frac{n-2}{16\pi r^2}\left(-\frac{2M(v)r^2}{(r^{n-2}+e^{n-2}(v))^{\frac{n-1}{n-2}}}+\frac{(n-1)M(v)r^{n}}{(r^{n-2}+e^{n-2}(v))^{\frac{2n-3}{n-2}}}-\frac{4\Lambda r^2}{(n-1)(n-2)}- \nonumber \right .
 \\& \left . (n-3)(\frac{M(v)r^2}{(r^{n-2}+e^{n-2}(v))^{\frac{n-1}{n-2}}}+ \frac{2\Lambda r^2}{(n-1)(n-2)})\right)
\end{align}  
	\section{Approaching the extremality and WEC}\label{sec5}
	In this section, we will study the third law of BH mechanics for the Vaidya metrics mentioned above. Recall that for these BHs, the Fodor and Hayward SGs are the same and are given by
	\begin{equation}\label{kappa}
		\kappa(v,r) =\frac{1}{2} f'(v,r)
	\end{equation}
	We follow the approach of section \ref{sec3} and assume that we don't know the exact location of the non-extremal BH horizons.
	\subsection{4D Vaidya BHs}
			It is convenient to consider the general form of \eqref{eq:L} in the case of $k\ne -\frac{1}{2}$ and set $\Lambda=0$. We can calculate the SG from  \eqref{kappa}
	\begin{align}
		\kappa(v,r) =\frac{m(v,r)}{r^2}-\frac{ m^{\prime}(v, r)}{r}=\frac{1}{r^2}\left(M(v)+\frac{8\pi k}{2k+1}C(v) r^{2k+1}\right)
	\end{align}
	to be evaluated at the AH. Consider a BH that is extremal at $v$ and its AH is at  $r_{ex}$, i.e.  $\kappa(v, r_{ex})=0$. This gives the radius of the AH as
		\begin{equation}\label{r11}
		r_{ex}^{2k+1}(v)=-\frac{(2k+1)M(v)}{8\pi k C(v)}
	\end{equation}
	Using the above with $f(v, r_{ex})=0$, we find 
	\begin{equation}\label{r21}
		r_{ex}^{2k}(v)=-\frac{1}{8\pi C(v)}
	\end{equation}
	and by comparing \eqref{r11} and \eqref{r21}
	\begin{equation}\label{r31}
		r_{ex}(v)=\frac{2k+1}{k}M(v)
	\end{equation}
	 From \eqref{r21} and \eqref{r31}, we see that $C(v)<0$ and the value of $k$ must be $k>0$ or $k<-1/2$ for $M(v)>0$. Thus, the extremality condition  is
	 	\begin{equation}\label{em}
		(\frac{1}{8\pi \rvert{C(v)\rvert}})^{\frac{1}{2k}}=\frac{(2k+1)M(v)}{k}
	\end{equation}
	which relates the parameters of the extremal BH at time $v$. Note that from \eqref{r31}, the radius of the extremal horizon is a simple root of $\kappa(v, r_{ex})=0$. 	
	
\begin{figure}[h]
\centering
\includegraphics[width=5in]{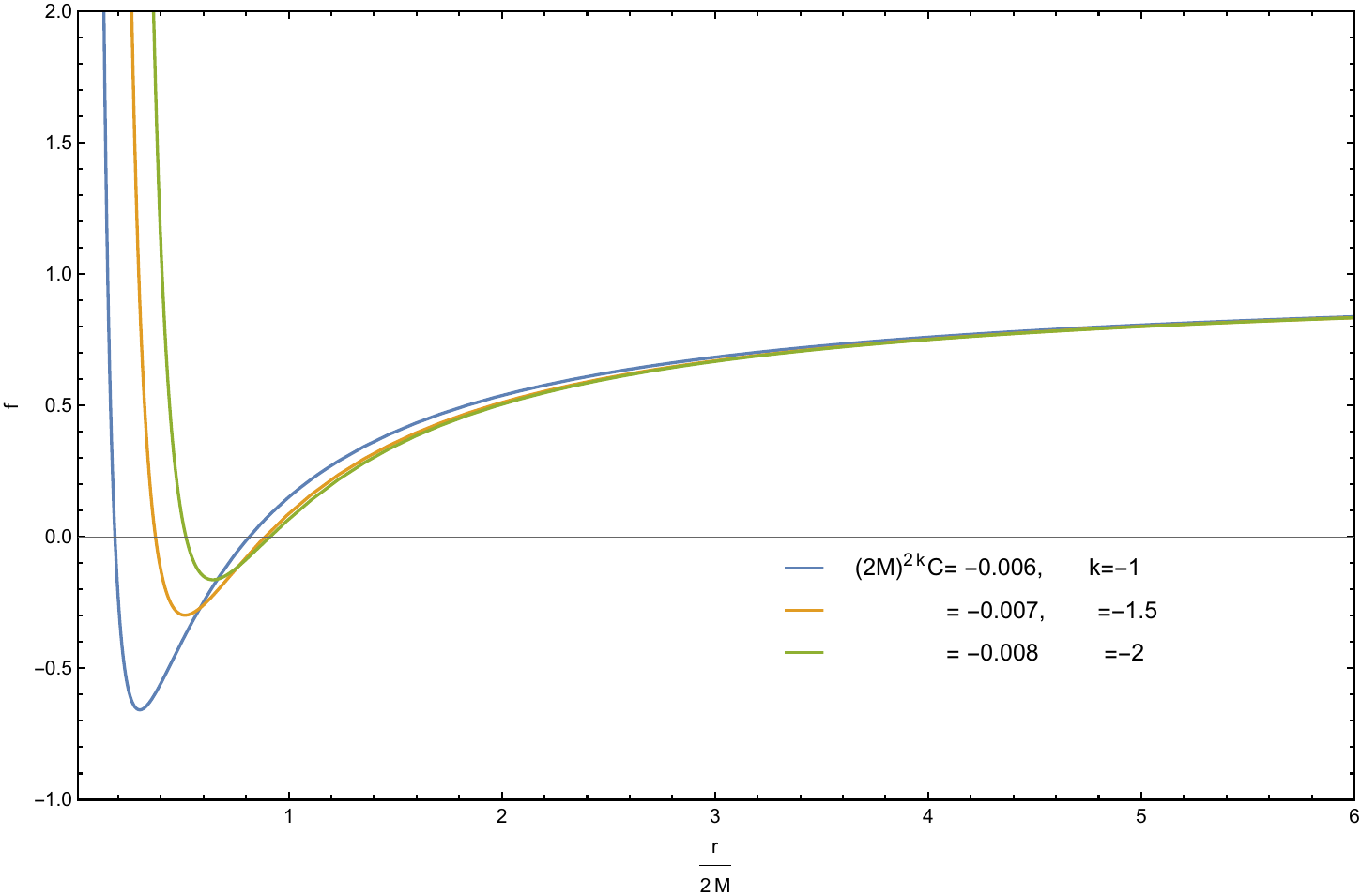}
\caption{The 4-D Vaidya metric function $f(v,r)$  with respect to the radius $r/2M$ for $k<-1/2$. Two horizons of the BH are represented by the intersection of the graph with the axis $r/2M$. }
\label{fig1}
\end{figure}

	For $k < -1/2$, the behavior of the function $f(v, r)$ at spatial infinity and $r \to 0$ shows that if the assumed BH is not extremal at $v$, then this function has two distinct roots according to Figure \eqref{fig1}, and a local minimum at $r_m$, where $r_m$ is the root of $f'(v, r)=0$ or $\kappa(v, r)=0$. 
Thus, the BH had at most two separate horizons at this time, and $f (v, r_{m})<0$. This condition, together with \eqref{r11}, gives 
	 	\begin{equation}\label{i}
		\frac{(2k+1)M(v)}{k} > (\frac{1}{8\pi \rvert C(v)\rvert})^{\frac{1}{2k}}
	\end{equation}
	for a non-extremal BH with $k < -1/2$. 
Note that the function $f(v, r)$ has two roots and a minimum with a negative value only if the values of $k$, $M(v)$ and $C(v)$ satisfy in \eqref{i}.
		Now, suppose that a dynamical non-extremal BH evolves and becomes extremal at $v_0$. This implies that $f (v, r_{m})<0$ for $v<v_0$ and  $f (v_0, r_{ex})=0$.
At $v_0$, the two horizons of the BH meet and $f (v_0, r)$ has a single-valued root at $r_{ex}$ where the extremality condition \eqref{em} is satisfied.  
To connect this situation with the WEC, it is convenient to introduce the function $\Delta$ which is defined as 
			 	\begin{equation}
	\Delta (v)\equiv \frac{(2k+1)M(v)}{k}-(\frac{1}{8\pi \rvert C(v)\rvert})^{\frac{1}{2k}}
		\end{equation}
		so that $\Delta(v_0)=0$.	
			 The BH is non-extremal before  $v=v_{0}$, hence $\Delta (v) > 0$ for $v<v_0 $. Looking
 at \eqref{i}, we see that $\Delta(v)$ is a decreasing function of advanced time as $v$ approaches to $v_0$. This requires that  
	\begin{equation}
		\dot{\Delta}\rvert_{v=v_{0}}=\frac{2k+1}{k}\left(\dot{M}(v_0)-\frac{M(v_0)\dot{C}(v_0)}{2k|C(v_0)|}\right)<0
	\end{equation}
	As mentioned above, $ (2k+1)/k$ is positive so
	\begin{equation}\label{r1}
		\dot{M}(v_0)-\frac{\dot{C}(v_0) M(v_0)}{2k\rvert C(v_0)\rvert} <0
	\end{equation}
	
\begin{figure}[h]
\centering
\includegraphics[width=5in]{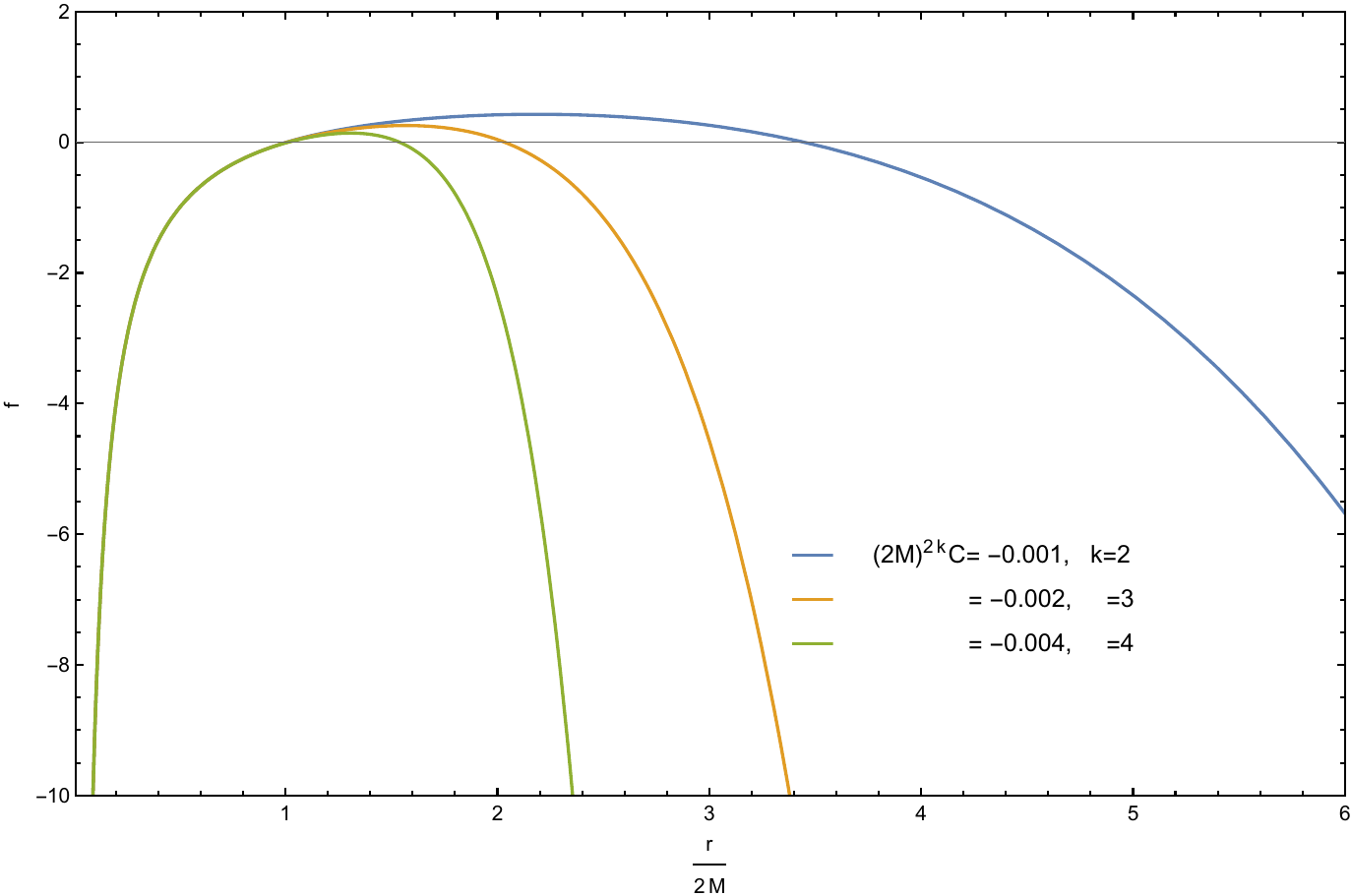}
\caption{The 4-D Vaidya metric function $f(v,r)$  with respect to the radius $r/2M$ for $k>0$. Two horizons of the BH are represented by the intersection of the graph with the axis $r/2M$.}
\label{fig2}
\end{figure}

For $k > 0$, according to Figure \eqref{fig2}, $f (v, r)$ has at most two separate horizons at $v < v_0$ and a local maximum at $r_m$, if the values of the constant $k$, the functions $M (v)$ and $C(v)$ satisfy the inequality $(2k+1)M(v)/k < (1/8\pi \rvert C(v)\rvert)^{1/2k}$. Thus, the sign of the inequality \eqref{i}, and consequently the sign of the inequality \eqref{r1}, changes when $f (v, r_{m})>0$. 

For an observer moving in the radial direction with four-velocity $u^{\alpha}$,  the WEC states
\begin{equation}
		\begin{split}
			T_{\alpha \beta} u^{\alpha} u^{\beta}
								&=g_{vv} T_v^v (u^v)^2 + g_{vr} T_r^r u^v u^r +g_{vr} T_v^r (u^v)^2+ g_{rv} T_v^v u^v u^r\\
			&=( -f(u^v)^2 + 2 u^v u^r) T^v_v+ T^r_v (u^v)^2 >0\\
		\end{split}
	\end{equation}
	which can be simplified to
		\begin{equation}\label{CDI}
		-T_v^v + T_v^r (u^v)^2 >0
	\end{equation}
	by noting that  the observer's velocity is normalized to	$ -f(u^v)^2 + 2 u^v u^r= -1$.
	Since the value of $(u^v)^2$ is arbitrary, the WEC \eqref{CDI} can be evaluated for sufficiently small and large values of  $(u^v)^2$. Thus, the WEC requires
	\begin{equation}\label{ine}
	T_v^v<0   \hspace{2cm}  T_v^r >0
	\end{equation}
 Considering \eqref{T12}, \eqref{hus22} and the two inequalities above at  $r_{ex}$, the first inequality is obviously satisfied by $C(v)<0$ and the other gives
	\begin{equation}
		\frac{1}{r_{ex}^2}(\frac{\dot{M}(v)}{4\pi }-\frac{\dot{C}(v) r^{2k+1}_{ex}}{2k+1}) >0
	\end{equation}
	which can be simplified to
	\begin{equation}\label{n}
		\dot{M}(v_0)-\frac{\dot{C}(v_0) M(v_0)}{2k\rvert C(v_0)\rvert} > 0
	\end{equation}
	at $v_0$ by using \eqref{r11}. 
	
\begin{figure}[h]
\centering
\includegraphics[width=5in]{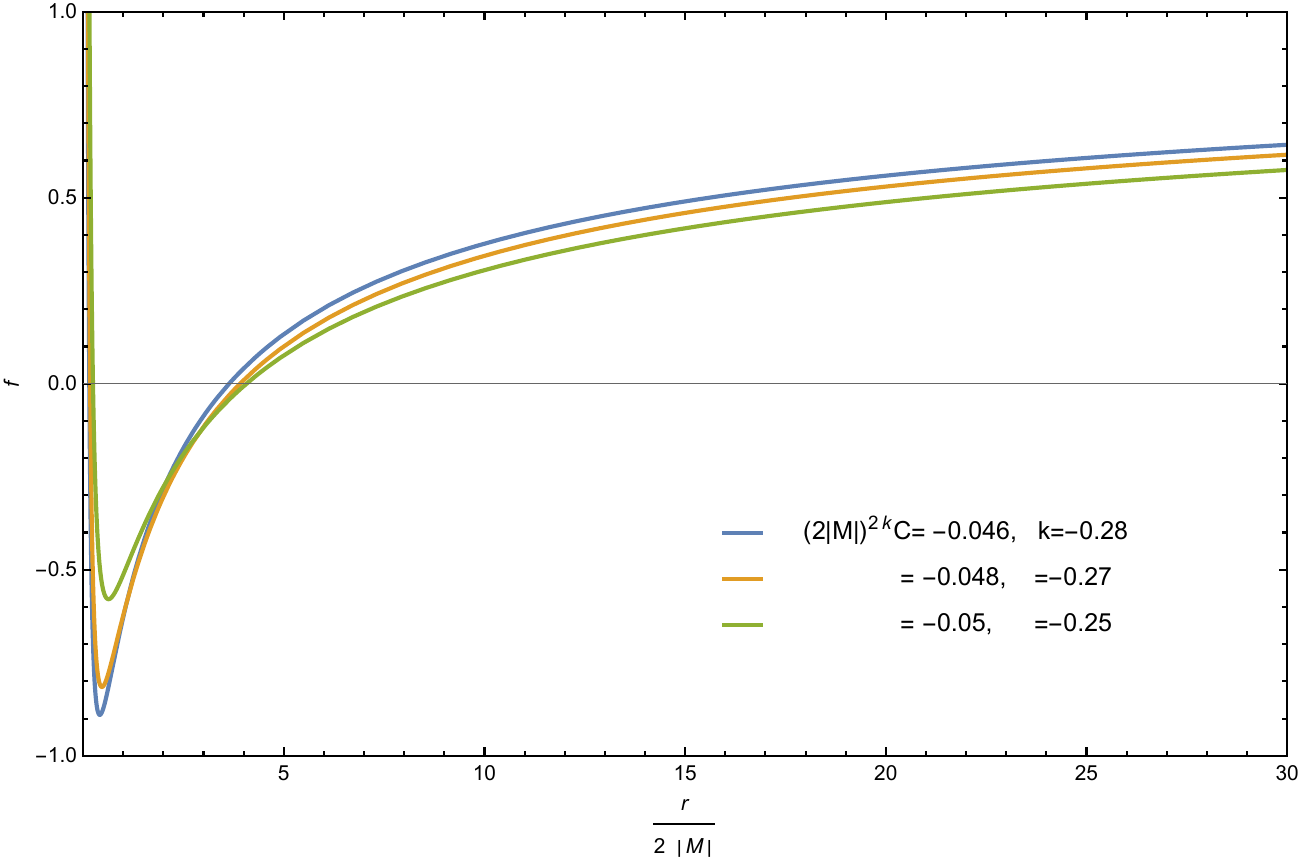}
\caption{The 4-D Vaidya metric function $f(v,r)$  with respect to the radius $r/2|M|$ for $-1/2<k<0$. Two horizons of the BH are represented by the intersection of the graph with the axis $r/2|M|$.}
\label{fig3}
\end{figure}

	Comparing \eqref{r1} with \eqref{n}, we see that the WEC prevents the BH from becoming extremal for  $k<-1/2$. However, for $k>0$ the condition required for the BH to reach extremality after finite time and the inequality we get from WEC are compatible. Note that this situation does not contradict what Israel states in \cite{Israel:1986gqz} since there is no trapped region in the region between two horizons. Israel associates the  extremalization of a BH with the “loss of its trapped surfaces”, which is a direct consequence of Raychaudhuri’s equation that trapped surfaces persist in evolution as long as the spacetime satisfies the WEC \cite{Kehle:2022uvc}. \\
	As mentioned in the \cite{Kothawala:2004fy}, the functions $M(v)$ and $C(v)$ are integration constants and it is not necessary that $M(v)$ is positive. Therefore, the parameter $k$ can be in the interval $(-1/2 , 0)$. It should be noted that, according to \eqref{eq:L}, the value of the mass function $m(v.r)$ can be positive or negative in both cases $M(v)>0$ or $M(v)<0$. For the latter, by examining the asymptotic behavior of the function $f(v,r)$ at spatial infinity and $r\to 0$, according to Figure \eqref{fig3}, we can see that, there is a trapped region between the two horizons. In this case, metric $(2)$ with the mass function $(10)$ has a local minimum and is asymptotically flat. Thus, with the condition $f (v, r_{m})<0$ for $v<v_0$, by assuming $-1/2<k<0$, the sign of inequality \eqref{i} and consequently the sign of inequality \eqref{r1} will change. In this case, the  extremality condition of the BH after a finite time and the inequality obtained from WEC are compatible. Thus, for $M(v)<0$ there is a clear violation of the third law of thermodynamics.
		\\
	Now we will  check the third law of BH mechanics for the general form of \eqref{eq:L} with $\Lambda=0$ in the case of $k=-\frac{1}{2}$. The corresponding SG would be
\begin{equation}\label{KK1}
\kappa(v,r)=\frac{M(v)}{r^2}-\frac{4\pi C(v)}{r^2}\ln(r)+\frac{4\pi C(v)}{r^2}
\end{equation}
The BH is extremal at $v$ if $\kappa(v,r_{ex})=0$, so that
\begin{equation}\label{rfg}
r_{ex}(v)=\exp(\frac{M(v)}{4\pi C(v)}+1)
\end{equation}
Using the above with $f(v, r_{ex}) = 0$ gives
\begin{equation}\label{reg1}
r_{ex}(v)=-8\pi C(v)
\end{equation}
From \eqref{reg1} we see that $C(v) < 0$. So the extremal condition is obtained
\begin{equation}\label{rl1}
M(v)=4\pi C(v)(\ln(8\pi |C(v)|)-1)
\end{equation}
Using $f (v, r_{m}) < 0$, we have
\begin{equation}\label{rll1}
\Delta (v)\equiv M(v)-4\pi C(v)(\ln(8\pi |C(v)|)-1)>0
\end{equation}
Thus, $\Delta (v)$ must decrease as $v$ approaches $v_{0}$.
\begin{equation}\label{r22}
\dot{\Delta}\rvert_{v=v_{0}}=\dot{M}(v_0) -4\pi \dot{C}(v_0) \ln(8\pi |C(v_0)|)<0
\end{equation}
Regarding the WEC, the first inequality of \eqref{ine} is satisfied by $C(v)<0$ and the evaluation of the second inequality at $r=r_{ex}$ gives 
\begin{equation}\label{r44}
\dot{M}(v_0)-4\pi \dot{C}(v_0)\ln(8\pi |C(v)|)>0
\end{equation}
Thus,  the WEC is a sufficient condition of the third law even for  $k=-1/2$. 	
	
	   Using  the above approach,  we have discussed the validity of the third law for some specific BHs. The results are summarized in Tables 1 and 2.
	 \begin{table*}[h!]
\caption{Some extremal Vaidya BHs}
\begin{center}
\renewcommand{\arraystretch}{1.1} 
\setlength{\tabcolsep}{0.2cm}
\label{table1}
\begin{minipage}{\linewidth}
\begin{tabular}{ |C{3cm}||C{3.9cm}|C{5cm}|C{5cm}| }
\hline
&&&\\[0.2mm]
BH& $\kappa=0$& $f(v,r_{ex})=0$ &  Extremality condition\\[7mm]
\hline
&&&\\[0.2mm]
Bonnor&   $r_{ex}(v)=\frac{q^{2}(v)}{M(v)}$  &  $r_{ex}(v)=q(v)$& $	M^2(v)=q^2(v)$\\[7mm]
\hline
&&&\\[0.2mm]
Husain\footnote{For this BH, the value of $r_{ex}$ shows that $(2m-1)/m$ is positive, so $m > 1/2$ or $m < 0$. This is used in Table 2 to compare the second and last column of the Husain BH.} & $ r_{ex}^{-2m+1}(v)=(\frac{2m-1}{2m}) \frac{M (v)}{g(v)}$& $r_{ex}^{-2m}(v)=\frac{1}{2g(v)}$&  $\frac{2m-1}{m} M(v)=(2g(v))^{\frac{1}{2m}}$ \\[7mm]
\hline
&&&\\[0.2mm]
Kiselev\footnote{For the Kiselev BH, the state parameter belongs to $-1<w<-1/3$} & $r_{ex}^{3w}(v)=-\frac{(1+3w)N(v)}{2M(v)}$ & $r_{ex}^{3w+1}(v)=-3wN(v)$ & $\frac{6w M(v)}{3w+1}=(-3wN(v))^\frac{1}{3w+1}$\\[5mm]
\hline
&&&\\[0.2mm]
A regular family\footnote{$\beta$ , $n>0$} & $r_{ex}^{-3/n}(v)=\frac{1}{2\beta(v)}$ & $r_{ex}(v)(1+\beta(v) r_{ex}^{-3/n}(v))=2m(v)$ & $\frac{2^{n+1}}{3^n}m(v)=(2\beta(v))^{\frac{n}{3}}$\\[5mm]
\hline
\end{tabular}
\end{minipage}
\end{center}
\end{table*}

\begin{table*}[h!]
\caption{Comparing inequalities from WEC and condition for evolving to future extremality}
\begin{center}
\renewcommand{\arraystretch}{1.1} 
\setlength{\tabcolsep}{0.2cm}
\label{table2}
\begin{tabular}{ |C{1.75cm}||C{3.9cm}|C{5.4cm}| C{1.5cm} | C{6.1cm} | }
\hline
&&&&\\[0.2mm]
BH&  $f (v, r_{m}) < 0$ &  $\dot{\Delta}(v_0)<0$ & $T^v_v<0$ & $T^r_v>0$ \\[7mm]\hline
&&&&\\[0.2mm]
Bonnor&   $\Delta(v)=M^2(v)-q^2(v)>0$  &  $M(v_0)\dot{M}(v_0)-q(v_0)\dot{q}(v_0) <0$& valid	& $M(v_0)\dot{M}(v_0)-q(v_0)\dot{q}(v_0)>0$\\[7mm]
\hline 
&&&&\\[0.2mm]
Husain & $ \Delta(v)= \frac{2m-1}{m}M(v)-(2g(v))^{\frac{1}{2m}}>0$ & $\frac{2m-1}{m} \dot M(v_0)-\frac{1}{m} \dot{g}(v_0)(2g(v_0))^{\frac{1}{2m} -1}<0$ &  valid & $\dot{M}(v_0)-(\frac{\dot{g}(v_0)}{2m-1})(2g(v_0))^{\frac{1}{2m}-1} >0$\\[7mm]
\hline
&&&&\\[0.2mm]
Kiselev &$\Delta(v)=\frac{6w M(v)}{3w+1} - (-3w N(v))^{\frac{1}{3w+1}}>0$ & $ \frac{3w}{3w+1}(2\dot{M}(v_0)+ \dot{N}(v_0)(-3wN(v_0))^{\frac{-3w}{3w+1}})<0$ &  valid & $2\dot{M}(v_0)+\dot{N}(v_0)(-3w N(v_0))^{\frac{-3w}{3w+1}}>0$ \\[5mm]
\hline
&&&&\\[0.2mm]
A regular family & $\Delta (v)=\frac{2^{n+1}}{3^n}m(v) -   (2\beta(v))^{\frac{n}{3}}>0$ & $ \frac{2^{n+1}}{3^n}(\dot{m}(v_0)-(\frac{n}{3})\frac{m(v_0)}{\beta(v_0)}\dot{\beta}(v_0))<0$ &  valid & $\dot{m}(v_0)-(\frac{n}{3})\frac{m(v_0)}{\beta(v_0)}\dot{\beta}(v_0)>0$ \\[5mm]
\hline
\end{tabular}
\end{center}
\end{table*} 
	\subsection{Extension to n-D Vaidya BHs}
\begin{figure}[h]
\centering
\includegraphics[width=5in]{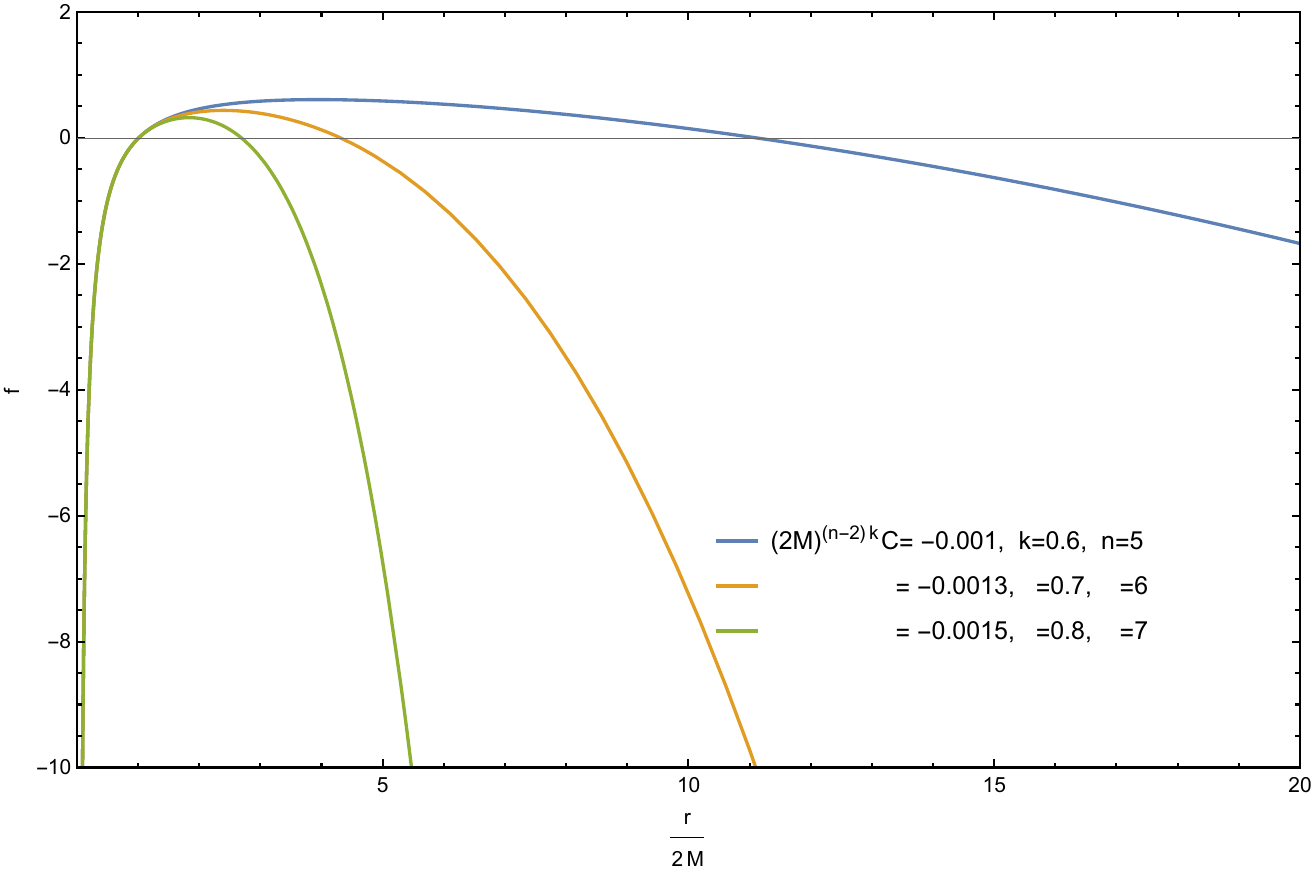}
\caption{The n-D Vaidya metric function $f(v,r)$  with respect to the radius $r/2M$ for $(n-2)k+1>n-3$. Two horizons of the BH are represented by the intersection of the graph with the axis $r/2M$.}
\label{fig4}
\end{figure}

\begin{figure}[h]
\centering
\includegraphics[width=5in]{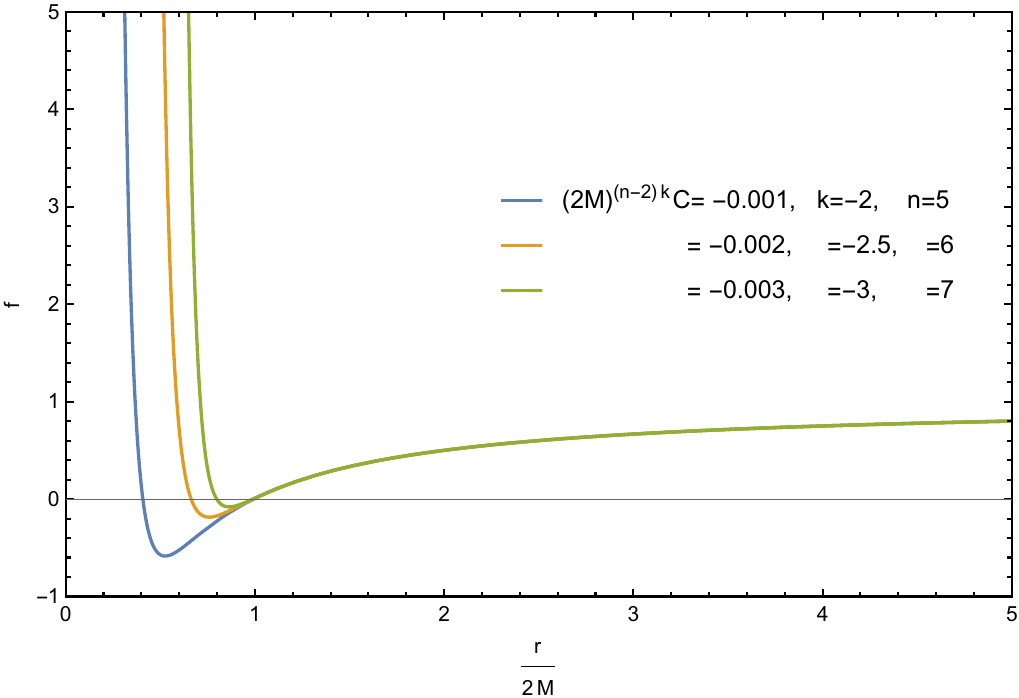}
\caption{The n-D Vaidya metric function $f(v,r)$  with respect to the radius $r/2M$ for $(n-2)k+1<0$. Two horizons of the BH are represented by the intersection of the graph with the axis $r/2M$.}
\label{fig5}
\end{figure}

\begin{figure}[h]
\centering
\includegraphics[width=5in]{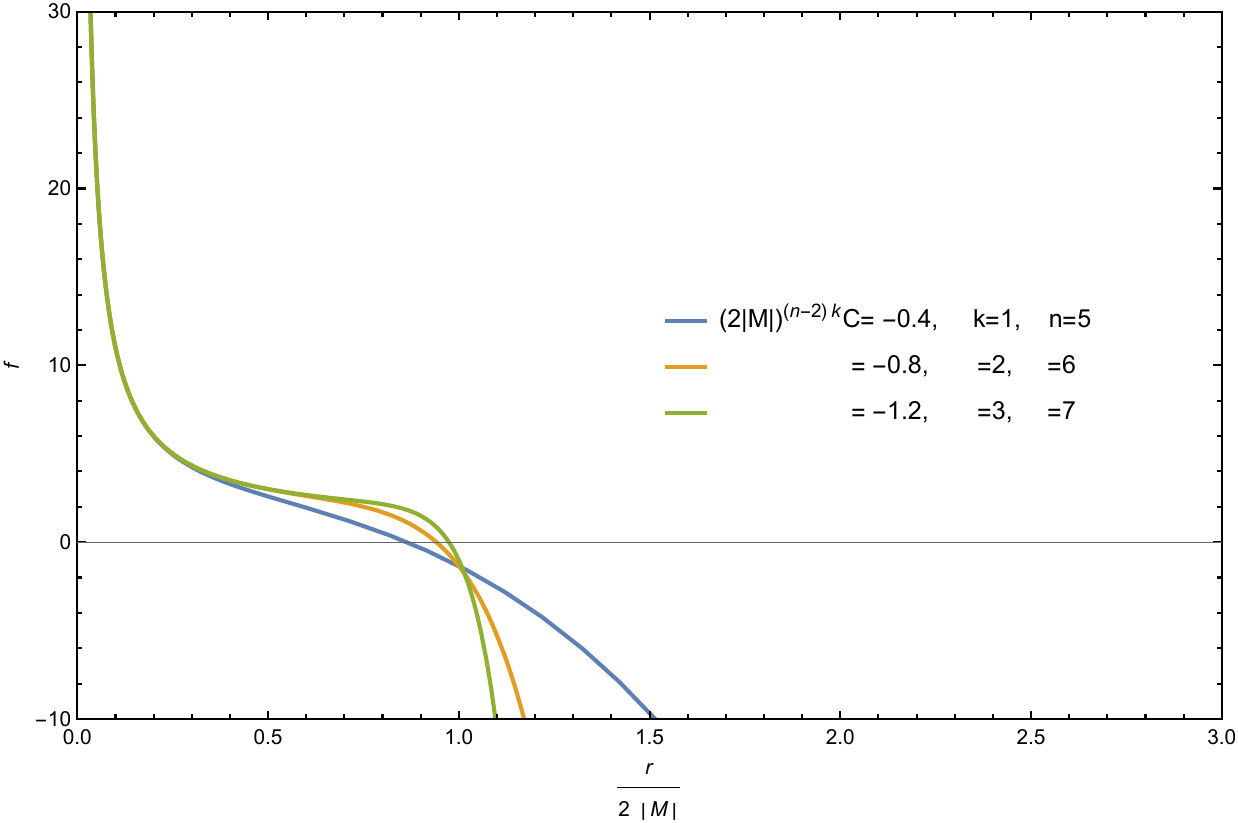}
\caption{The n-D Vaidya metric function $f(v,r)$  with respect to the radius $r/2|M|$ for $(n-2)k>0$ Two horizons of the BH are represented by the intersection of the graph with the axis $r/2|M|$.}
\label{fig6}
\end{figure}

\begin{figure}[h]
\centering
\includegraphics[width=5in]{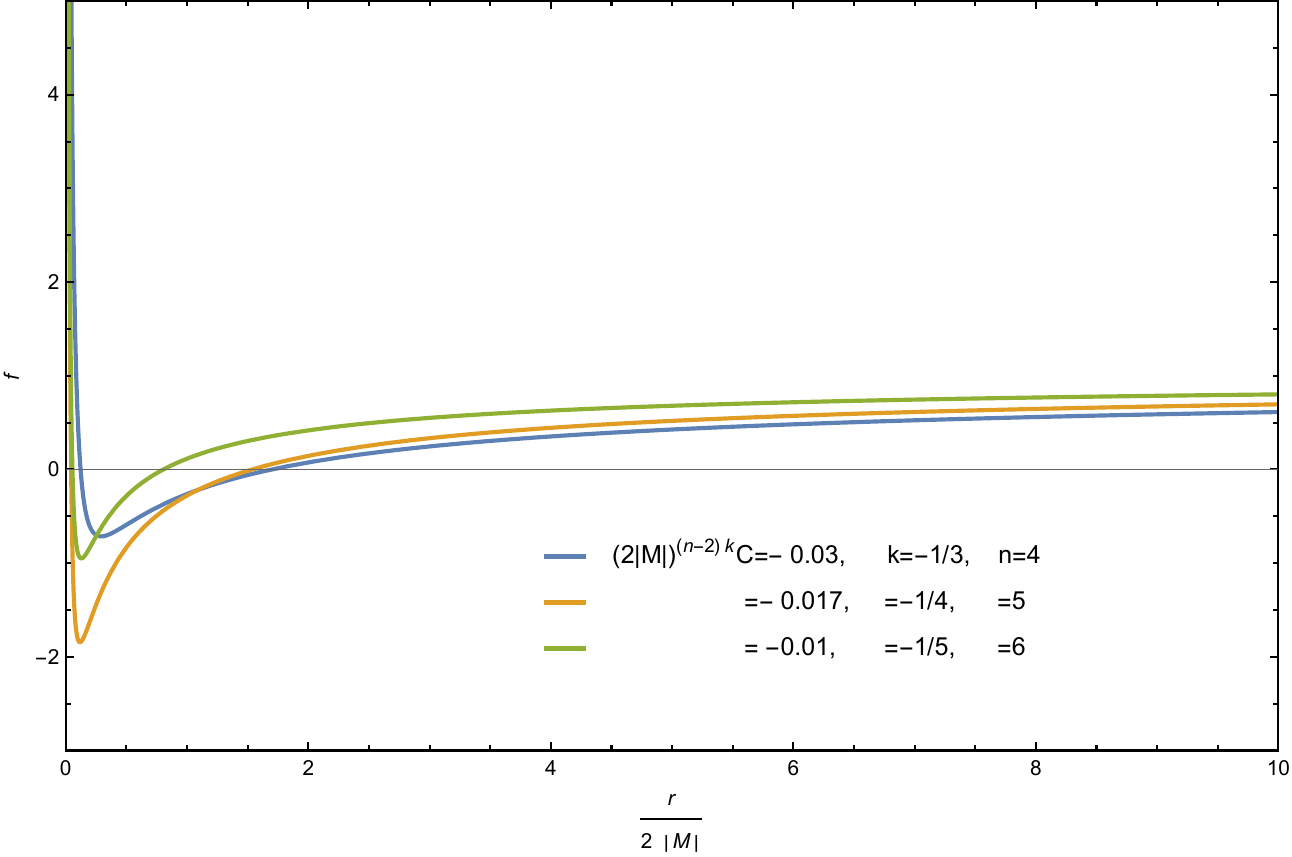}
\caption{The n-D Vaidya metric function $f(v,r)$  with respect to the radius $r/2|M|$ for $-1<(n-2)k<0$. Two horizons of the BH are represented by the intersection of the graph with the axis $r/2|M|$.}
\label{fig7}
\end{figure}

	The extension of our analysis to n-D BHs is straightforward. We consider the family of n-D Vaidya BHs \eqref{dn8}, with $\Lambda=0$ for the case that $k \ne -1/(n-2)$. The SG of the BH is
		\begin{equation}\label{rg2}
		\begin{split}
			 \kappa(v,r)=\frac{1}{r^{n-2}}\left(M(v) -\frac{8 \pi C(v)}{n-2} r^{(n-2)k+1}\left(1-\frac{n-3}{(n-2)k+1}\right)
			   \right) 
		\end{split}	
	\end{equation}
From the fact that $\kappa(v, r_{ex}) = 0$ for the extremal BH, we get
	\begin{equation}\label{rg5}
		r_{ex}^{(n-2)k+1}(v) = -\frac{ M(v) (n-2) }{ 8\pi  C(v)}  \left(1-\frac{n-3}{(n-2)k+1}\right)^{-1}
	\end{equation}
	showing  that $r_{ex}$ is a function of both $M (v)$ and $C(v)$. Substituting $M (v)$ from \eqref{rg5} into  $f (v, r_{ex}) = 0$ yields
	\begin{equation}\label{rg7}
		r_{ex}^{(n-4)-(n-2)k}(v) =- \frac{ 16\pi  C(v) } {(n-3)(n-2)} 
	\end{equation}
	So, $C(v)<0$ and by assuming $M(v)>0$, it should be $(n-2)k+1>n-3$ or $(n-2)k+1<0$. By comparing \eqref{rg5} and \eqref{rg7}
		\begin{equation}\label{rg8}
		r_{ex}^{n-3}(v) = \frac{2 M(v)}{n-3} \left(1-\frac{n-3}{(n-2)k+1}\right)^{-1}
	\end{equation}
	 and the extremality condition is obtained as follows
	 	 \begin{equation}\label{rg9}
		\left(\frac{n-2}{8\pi |C(v)|}\right)^{n-3}\left(\frac{n-3}{2}\right)^{(n-2)k+1}=\left(M(v)\left(1-\frac{n-3}{(n-2)k+1}\right)^{-1}\right)^{(n-2)k-(n-4)}
	\end{equation}
According to the behavior of the function $f(v,r)$ plotted in figures \eqref{fig4} and \eqref{fig5} for $(n-2)k+1>n-3$ and $(n-2)k+1<0$ respectively, we can see that in the first case there is no trapped region between two horizons. In this case, similar to what was mentioned in the previous section in 4D, the inequality resulting from the WEC is compatible with the inequality resulting from the BH extremality condition during a finite time process. However, this situation is not our concern since it does not contradict the third law. In the case $(n-2)k+1<0$, for a non-extremal BH which would be extremal at $v_0$ , the condition $f (v, r_{m}) < 0$ for $v < v
	_0$ implies  
		\begin{equation}\label{rg11}
		\Delta(v) =\left(M(v)\left(1-\frac{n-3}{(n-2)k+1}\right)^{-1}\right)^{(n-2)k-(n-4)}-\left(\frac{n-2}{8\pi |C(v)|}\right)^{n-3}\left(\frac{n-3}{2}\right)^{(n-2)k+1}<0
	\end{equation}
	Since $\Delta(v)$ increases as $v$ approaches $v_0$, thus
	\begin{equation}\label{rg16}
		\dot{\Delta} \rvert_{v=v_{0}}=r_{ex}^{(n-3)\left((n-2)k-(n-4)\right)}\left(\frac{n-3}{2}\right)^{(n-2)k-(n-4)}\left(\left((n-2)k-(n-4)\right)\frac{\dot{M}(v)}{M(v)}-(n-3)\frac{\dot{C}(v)}{|C(v)|}\right)>0
	\end{equation}
which can be simplified to 
\begin{equation}\label{rg17}
		\frac{1}{n-3}\frac{\dot{M}(v_0)}{M(v_0)}- \frac{1}{(n-2)k-(n-4)} \frac{\dot{C}(v_0)}{\rvert C(v_0) \rvert} <0
	\end{equation}
since $ (n-2)k-(n-4)<0$ and $n>3$.	
Given \eqref{dn6} and \eqref{dn10}, a simple calculation shows that since $C(v)<0$, the first condition of WEC \eqref{ine} is satisfied  and the second condition gives 
	\begin{equation}\label{rg18}
		T_v^r = \frac{1}{8\pi} r_{ex}^{-(n-2)} (\frac{n-2}{n-3}) \dot{M}(v_0 )-\frac{1}{(n-2)k+1} \dot{C}(v_0) r_{ex}^{(n-2)k-(n-3)} >0
	\end{equation}
	 With $n > 3$ and using \eqref{rg5}, the result is
	\begin{equation}\label{rg19}
		\frac{1}{8\pi}\left(\frac{n-2}{n-3} \dot{M}(v_0)- \frac{M(v_0) (n-2)}{(n-2)k-(n-4)} \frac{\dot{C}(v_0)}{\rvert C(v_0) \rvert}\right) >0
	\end{equation}
	Since  $8\pi/(M(v) (n-2)) >0$, the above inequality is in contradiction with \eqref{rg16}. Therefore for $\Lambda=0$, $k \ne -1/(n-2)$ and $n>3$, the family of n-D  radiating BHs \eqref{dn8} obeying the  WEC, cannot become extremal at finite time $v_0$. \\
	In the situation where $M(v)<0$, it must be $(n-2)k+1>0$ and $(n-2)k+1<n-3$, so that the right side of \eqref{rg5} is positive. For  dimensions greater than 4, $(n-2)k$ can also be in a positive interval because $-1<(n-2)k<n-4$ where $n-4>0$. 
If $(n-2)k>0$, according to the behavior of the function $f(v,r)$ shown in Figure \eqref{fig6} , it is possible that the BH before time $v_0$ has only one horizon (it is the event horizon), which is not our desired situation.
If $(n-2)k$ is in the interval $(-1, 0)$, then according to the behavior of the function $f(v,r)$ plotted in Figure \eqref{fig7}, we will have a trapped region between the two horizons, and due to the change of the sign of inequality \eqref{rg17} and its compatibility with the inequality resulting from satisfying the WEC, the third law of thermodynamics will be violated.

Due to the fact that $\kappa(v, r_{ex}) = 0$ for the extremal BH,  when $k = -1/(n - 2)$ we get
	\begin{equation}\label{tob2}
	\ln(r_{ex})=\frac{1}{n-3}(1+\frac{(n-2)M(v)}{8\pi C(v)})
	\end{equation}
	Using the above with $f(v, r_{ex}) = 0$, we get
	\begin{equation}\label{tab3}
r_{ex}^{n-3}(v)=\frac{-16\pi C(v)}{(n-2)(n-3)}
	\end{equation}
	 Thus, $C(v)<0$. Comparing \eqref{tob2} and \eqref{tab3}, we have the extremal condition
	\begin{equation}\label{tab4}
M(v)=\frac{8\pi C(v)}{n-2}\left((n-3)\ln\left(\frac{16\pi |C(v)|}{(n-2)(n-3)}\right)^{\frac{1}{n-3}}-1\right)
	\end{equation}
	Since $f (v, r_{m}) < 0$, we have
	\begin{equation}\label{tab5}
M(v)>\frac{8\pi C(v)}{n-2}\left((n-3)\ln\left(\frac{16\pi |C(v)|}{(n-2)(n-3)}\right)^{\frac{1}{n-3}}-1\right)
	\end{equation}
		For a BH that becomes extremal at a finite time $v_{0}$ we have $\Delta (v_{0})= 0$ where 
	\begin{equation}
	\Delta (v)\equiv M(v)-\frac{8\pi C(v)}{n-2}\left((n-3)\ln\left(\frac{16\pi |C(v)|}{(n-2)(n-3)}\right)^{\frac{1}{n-3}}-1\right)
	\end{equation}
		 The BH is non-extremal before  $v=v_{0}$, so $\Delta (v) > 0$ for $v<v_0$ and thus, $\Delta (v)$ must decrease as $v$ approaches $v_{0}$. This yields
	\begin{equation}
		\dot{\Delta}\rvert_{v=v_{0}}=\dot{M}(v_0)-\frac{8\pi(n-3) \dot{C}(v_0)}{n-2}\ln\left(\frac{16\pi |C(v_0)|}{(n-2)(n-3)}\right)^{\frac{1}{n-3}}<0
	\end{equation}
	As for the WEC, the first inequality of \eqref{ine} is satisfied by $C(v)<0$ and evaluating the second inequality at $r=r_{ex}$ for $n>3$ gives 
\begin{equation}\label{tob8}
\dot{M}(v_0)-\frac{8\pi(n-3) \dot{C}(v_0)}{n-2}\ln\left(\frac{16\pi |C(v_0)|}{(n-2)(n-3)}\right)^{\frac{1}{n-3}}>0
\end{equation}
	Thus, the family of n-D radiating BHs \eqref{dn8} with $k=-1/(n-2)$ satisfying the WEC cannot become extremal in finite time.
	
It is straightforward to perform similar calculations for n-D uncharged de Sitter (metric \eqref{ADQ} with $q=0$) and for the n-D Bardeen BHs (metric  \eqref{Bar} with $\Lambda=0$).  The results are shown in Tables \ref{table 3} and \ref{table4}.
	 \begin{table*}[h!]
\caption{Some extremal n-D Vaidya BHs} 
\begin{center}
\renewcommand{\arraystretch}{1.1} 
\setlength{\tabcolsep}{0.2cm}
\label{table 3}
\begin{tabular}{ |C{3cm}||C{3.9cm}|C{5cm}|C{5cm}| }
\hline
&&&\\[0.2mm]
n-D BH& $\kappa=0$& $f(v,r_{ex})=0$ &  Extremality condition\\[7mm]
\hline
&&&\\[0.2mm]
Uncharged de Sitter&   $r_{ex}(v)=\left(m(v) l^2 (n-3)\right)^\frac{1}{n-1}$  &  $r_{ex}(v)=\sqrt{\frac{(n-3)l^2}{n-1}}$ & $	m(v)(n-1)=\left(\frac{(n-3)(n-2)}{2\Lambda}\right)^{\frac{n-3}{2}}$\\[7mm]
\hline
&&&\\[0.2mm]
Bardeen&   $r_{ex}(v)=(\frac{2e^{n-2}(v)}{n-3})^{\frac{1}{n-2}}$  &  $r_{ex}(v)=\left((\frac{2}{n-1})^{\frac{n-1}{n-2}}M(v)\right)^{\frac{1}{n-3}}$ & $	M(v)=\left(\frac{(n-1)^{n-1}(n-3)^{3-n}}{4}\right)^{\frac{1}{n-2}}e^{n-3}(v) $\\[7mm]
\hline
\end{tabular}
\end{center}
\end{table*}
\begin{table*}[h!]
\caption{Comparing inequalities from WEC and condition for evolving to future extremality}
\begin{center}
\renewcommand{\arraystretch}{1.1} 
\setlength{\tabcolsep}{0.2cm}
\label{table4}
\begin{tabular}{ |C{1.9cm}||C{4.1cm}|C{5cm}| C{1.5cm} | C{5.4cm} | }
\hline
&&&&\\[0.2mm]
BH&  $f (v, r_{m}) < 0$ &  $\dot{\Delta}(v_0)<0$ & $T^v_v<0$ & $T^r_v>0$ \\[7mm]\hline
&&&&\\[0.2mm]
Uncharged de-Sitter&   \makecell{\vspace{0.1cm}$\Delta(v)=  	m(v)(n-1)-$ \\ $ \left(\frac{(n-3)(n-2)}{2\Lambda}\right)^{\frac{n-3}{2}}>0$\\ $ $}  &  $\dot{m}(v_0)<0$& valid	& $\dot{m}(v_0)>0$\\[7mm]
\hline 
&&&&\\[0.2mm]
Bardeen&   \makecell{\vspace{0.1cm}$\Delta(v)=M(v)-$ \\ \vspace{0.2cm} $\left(\frac{(n-1)^{n-1}(n-3)^{3-n}}{4}\right)^{\frac{1}{n-2}}$ \\ $ \times e(v)^{n-3}>0$ \\ $ $}  &  \makecell{\vspace{0.1cm}$\dot{M}(v_0)-\left(\frac{(n-3)(n-1)^{n-1}}{4}\right)^{\frac{1}{n-2}}$ \\ $\times e^{n-4}(v_0)\dot{e}(v_0)<0$} & valid	& \makecell{\vspace{0.1cm}$\dot{M}(v_0)-\left(\frac{(n-3)(n-1)^{n-1}}{4}\right)^{\frac{1}{n-2}}$ \\ $ \times e^{n-4}(v_0)\dot{e}(v_0)>0$}\\[7mm]
\hline 
\end{tabular}
\end{center}
\end{table*} 
\section{Multi-Horizon BHs}\label{secpre6}
\begin{figure}[h]
\centering
\includegraphics[width=4in]{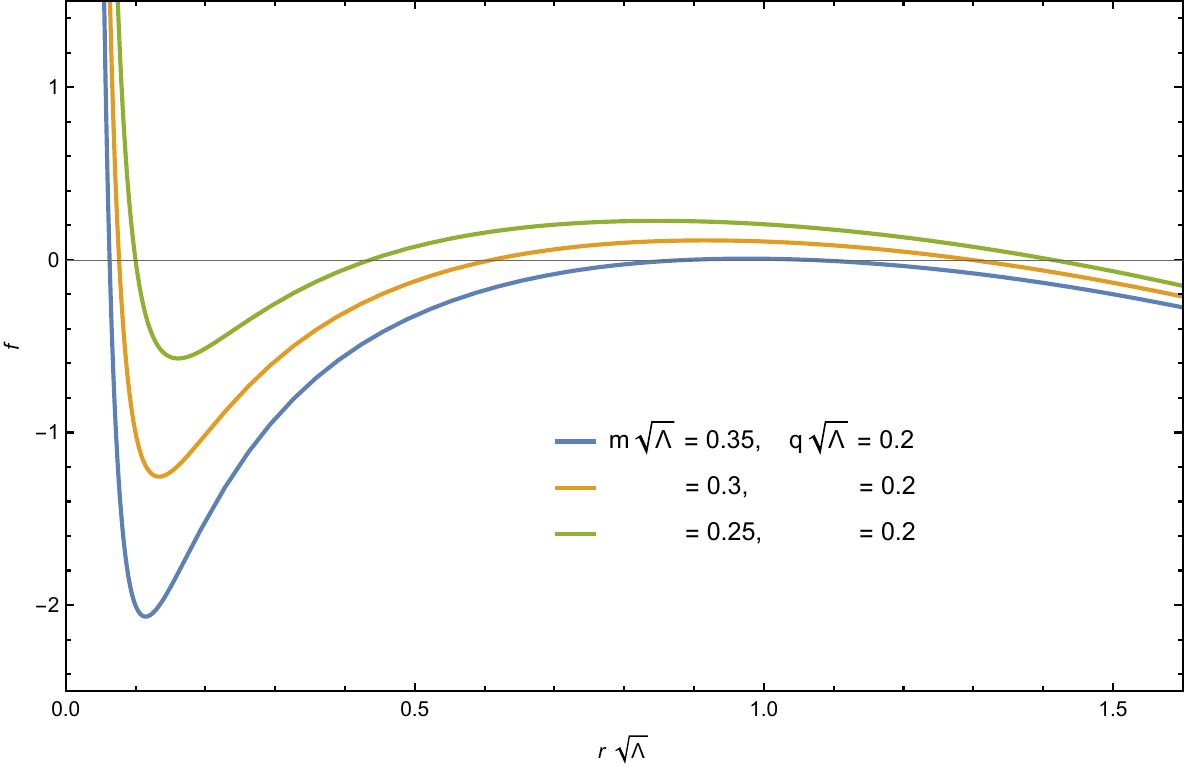}
\caption{The function $f(v,r)$ of the 4-D Vaidya de Sitter-Reissner-Nordstr\"om metric with respect to the dimensionless radius $\sqrt{\Lambda}r$.The intersection of the graph with the axis $\sqrt{\Lambda}r$ represents two horizons of the BH with radii $r_1$ and $r_2$ and the cosmological AH, with radius $r_3$, respectively.}
\label{fig}
\end{figure}

In this section, we will consider the generalization of the proposed method to multi-horizon BHs. For this purpose, we consider the Vaidya de Sitter-Reissner-Nordstr\"om metric \eqref{ADQ} in 4-D, which has three horizons. The outermost horizon, $r_3$, is the cosmological AH. By definition, the cosmological AH is obtained by setting the expansion parameter of the ingoing null geodesics, $\theta_{n}$, to zero while $\theta_{\ell}$ is positive \cite{faraoni}. According to Figure (\ref{fig}), there is no trapped region between horizon $r_3$ and $r_2$. The two inner horizons represent the two horizons of the charged Reissner-Nordstr\"om BH in a de Sitter background. 
Here we study the extremality condition for two outer horizons. If these coincide at radius $r_{ex}$ in finite advanced time $v_0$,  the conditions $f(v, r_{ex}) = 0$ and $f'(v, r_{ex}) = 0$  lead to the following equations, respectively:
\begin{align}
-\Lambda r^4_{ex}+3r^2_{ex}-6m(v)r_{ex}+3q^2(v)&=0 \label{dosi1}\\
\Lambda r^4_{ex}-3m(v)r_{ex}+3q^2(v)&=0\label{dosi2}
\end{align}
and by combining them, we get
\begin{align}
r_{ex}^2-3m(v)r_{ex}+2q^2(v)&=0 \label{dosi3}\\
\Lambda r^4_{ex}-r^2_{ex}+q^2(v)&=0\label{dosi4}.
\end{align} 
We have $f(v, r_{m})>0$ for $v < v_0$ at a local maximum $r_m$, which is the root of $f'(v, r_{m})$. This condition for $v < v_0$ leads to $r^2_{m}(v) f(v, r_{m})=r^2_{m}(v)-3m(v)r_{m}(v)+2q^2(v)>0$.
Now, suppose that a dynamical non-extremal BH evolves and becomes extremal at $v_0$.
Then it is convenient to introduce the function $\Delta$ which is defined as $\Delta(v)\equiv f(v, r_{m})$. Thus,  $\Delta(v_0)=f(v_0, r_{ex})=0$ and $\Delta(v)$ is a decreasing function of advanced time as $v$ approaches to $v_0$:
\begin{align}
\dot {\Delta}(v)=\dot{r}_{m}(v)\left(2r_{m}(v)-3m(v)\right)-3\dot{m}(v)r_{m}(v)+4q(v)\dot{q}(v)<0\label{dosi5}.
\end{align} 
Evaluating it at time $v_0$ by using \eqref{dosi3},  we can rewrite $2r_{m}(v)-3m(v)$ in terms of $r_{ex}-2q^2(v_0)/r_{ex}$. Also by differentiating $f(v, r_{m})=0$  with respect to the advanced time $v$ and using \eqref{dosi4}, we can simplify \eqref{dosi5} to
\begin{align}
\dot{\Delta}\rvert_{v=v_{0}}=\dot{m}(v_0)r_{ex}-q(v_0)\dot{q}(v_0)>0\label{dosi6}.
\end{align} 
Regarding the WEC, the first inequality of \eqref{ine} is satisfied according to \eqref{poi} and the evaluation of the second inequality for $r=r_{ex}$ gives
\begin{align}
\dot{m}(v_0)r_{ex}-q(v_0)\dot{q}(v_0)>0\label{dosi7}
\end{align} 
which is compatible with \eqref{dosi6}. If we do the same calculations for the two inner horizons, since at $v < v_0$ the value of $f(v, r_{m})$ is negative, the sign of the inequality \eqref{dosi6} changes and WEC is a sufficient condition of the third law. Thus, two inner horizons of the BH cannot become extremal in finite time, but the cosmological AH can coincide with the outer horizon of the BH in finite time without violating the WEC.
\section{Conclusion}\label{sec6}
In this paper, the third law of thermodynamics is studied for the families of 4-D  and n-D  Vaidya BHs, taking into account that for the non-stationary BHs, a quasi-local horizon is used instead of the Killing horizon, and also, there are different prescriptions for the SG of a non-Killing horizon.
We have used the Fodor and Hayward definitions of SG which are equivalent for all metrics considered here. The other definitions are not considered here because they do not give the correct value of SG in stationary space-time or they use a special normalization of the outgoing null vector.
We have shown that the coordinate invariant definition of extremality of a BH corresponds to zero SG for the evolving  horizons considered here. 
This is similar to what is expected for stationary extremal BHs, where the Killing SG should be zero. According to the third law, the SG of a BH cannot be reduced to zero in finite advanced time if the stress-energy tensor of the matter satisfies the WEC. In order to test this law, we assume that the non extremal BH evolves  to the extremal one, i.e. zero SG, in finite time. Then, we have compared the result with the WEC condition. In this sense, we have investigated whether we can obtain an extremal BH by  adjusting the free parameters of the metric.

Our results show that the WEC prevents the BH from becoming extremal at finite advanced time in a continuous process for a large family of 4-D and n-D Vaydia BHs. However,
assuming a negative integration constant $M(v)$, appearing in the mass function $m(v,r)$, it is possible to violate the third law of thermodynamics for certain intervals of $k$.
In other words, while the energy tensor of matter remains bounded and satisfies the WEC in a neighborhood of the AH, we will be able to remove the trapped region between two horizons during an appropriate dynamical process in finite time. Therefore, the third law of thermodynamics of BHs is indeed violated.
There are also some values of the metric parameters or in some cases such as the BHs with multiple horizons, where there is no trapped region between two horizons. Our calculations show that these horizons can coincide in a finite advanced time while the WEC is satisfied. However, due to the absence of the trapped region, such cases are not considered as counterexamples of the formulation of Israel’s third law.
\section*{Acknowledgement}
We would like to thank the anonymous referee for his/her valuable comments. F.B and F.S would like to thank the Iran National Science Foundation (INSF) for supporting this research under grant
number 4021095. F.S. is grateful to the University of Tehran for supporting this work under
a grant provided by the University Research Council.

\end{document}